# Simplest AB-Thermonuclear Space Propulsion and Power Source*

**Alexander Bolonkin**
C&R, 1310 Avenue R, #F-6, Brooklyn, NY 11229, USA
T/F 718-339-4563, aBolonkin@juno.com, http://Bolonkin.narod.ru

## Abstract

The author applies, develops and researches mini-sized Micro- AB Thermonuclear Reactors for space propulsion and space power systems. These small engines directly convert the high speed charged particles produced in the thermonuclear reactor into vehicle thrust or vehicle electricity with maximum efficiency. The simplest AB-thermonuclear propulsion offered allows spaceships to reach speeds of 20,000 - 50,000 km/s (1/6 of light speed) for fuel ratio 0.1 and produces a huge amount of useful electric energy. Offered propulsion system permits flight to any planet of our Solar system in short time and to the nearest non-Sun stars by E-being or intellectual robots during a single human life period.

**Key words:** AB-propulsion, thermonuclear propulsion, space propulsion, thermonuclear power system.

## Introduction

At present, both solid and liquid chemical fueled rockets are used for launch to and flights in interplanetary outer space. They have been intensively developed since War II when German engineer Wernher von Braun (1912-1977) successfully designed the first long distance rocket FAU-2. In the subsequent years, liquid and solid rockets reached their developmental peak. Their main shortcomings are (1) very high space launch cost of $20,000 – 50,000/kg; (2) large fuel consumption; (3) liquid fuel storage problems because oxidizer and fuel (for example; oxygen and hydrogen) require cryogenic temperatures, or they are poisonous substances (for example; nitric acid, $N_2O_3$).

In past years, the author and other scientists have published series of new methods that promise to revolutionize space launch and space flight [1-14]. These include cable accelerator, circle launcher and space keeper, space elevator transport system, space towers, kinetic towers, the gas-tube method, sling rotary method, electromagnetic and electrostatic accelerators, tether system, Earth-Moon or Earth-Mars non-rocket cable transport system. There include new propulsion and power systems such as solar and magnetic sails, Solar wind sail, radioisotope sail, electrostatic space sail, laser beam, kinetic anti-gravitator, multi-reflective beam propulsion system, asteroid employment electrostatic levitation, etc. (Too, there are new ideas in aviation that can be useful for flights in the atmosphere.)

Some of these have the potential to decrease space launch costs thousands of times, others allow changing the speed and direction of space apparatus without expending fuel.

The thermonuclear propulsion and power method is very perspective -- though not speculative -- because it promises high vehicular apparatus speed up 50,000 km/s. This method needs a small, special thermonuclear reactor that will allow the direct and efficient utilization of the kinetic energy of nuclear particles – the AB Thermonuclear Reactor –first offered by author [15].

## Description of innovations

The AB thermonuclear propulsion and electric generator are presented in fig.1. As it is shown in [15] the minimized, or micro-thermonuclear reactor 1 generates high-speed charged particles 2 and neutrons that leave the reactor. The emitted charged particles may be reflected by electrostatic reflector, 4, or adsorbed by a semi-spherical screen 3; the neutrons may only be adsorbed by screen 3.

In *screen* of the AB-thermonuclear reactor (fig.1*a*) the forward semi-spherical screen 3 adsorbed particles that move forward. The particles, 2, of the back semi-sphere move freely and produce the vehicle's thrust. The forwarded particles may to warm one side of the screen (the other side is heat protected) and emit photons that then create additional thrust for the apparatus. That is the *photon* AB-thermonuclear thruster.

In *reflector* AB-thermonuclear reactor (fig.1*b*) the neutrons fly to space, the charged particles 5 are reflected the electrostatic reflector 4 to the side opposed an apparatus moving and create thrust.

The *screen-reflector* AB-thermonuclear reactor (fig.1*c*) has the screen and reflector

The *spherical* AB-propulsion-generator (fig.1*d*) has two nets which stop the charged particles and produced electricity same as in [14] Chapter 17. Any part 8 of the sphere may be cut-off from voltage and particles 9 can leave the sphere through this section and, thusly, create the thrust. We can change direction of thrust without turning the whole apparatus.

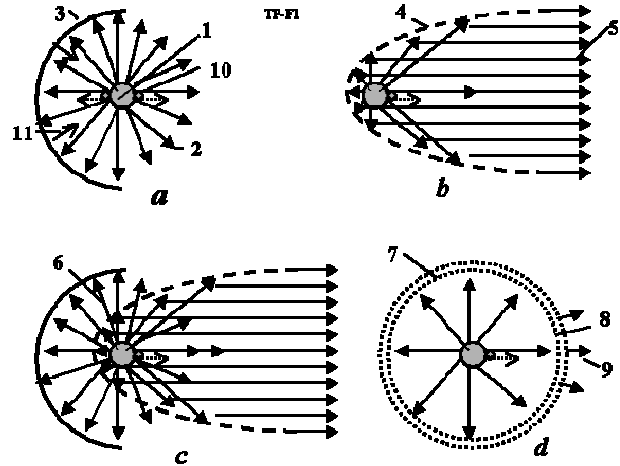

**Fig.1**. Types of the suggested propulsion and power system. (*a*) screen AB-thermonuclear propulsion and *photon* AB-thermonuclear propulsion ; (*b*) (electrostatic) *reflector* AB-thermonuclear propulsion; (*c*) *screen-reflector* AB-thermonuclear propulsion; (*d*) *spherical* AB-propulsion-generator. Notations: 1 - micro (mini) AB-thermonuclear reactor [15], 2 - particles (charged particles and neutrons) , 3 - screen for particles, 4 - electrostatic reflector; 5 - charged particles, 6 - neutrons, 7 - spherical net of electric generator, 8 - transparency (for charged particles) part of spherical net, 9 - charged particle are producing the thrust, 10 - electron discharger, 11 - photon radiation.

## Theory of the thermonuclear reactor, propulsion and power
### List of main equations

Below are the main equations for the proper estimation of benefits from the offered innovations.
**1. Energy needed to overcome the Coulomb barrier**

$$F = k\frac{Q_1 Q_2}{r^2}, \quad E = \int_{r_0}^{\infty} F dr, \quad E = \frac{kZ_1 Z_2 e^2}{r_0}, \quad (1)$$

$$r_0 = (1.2 \div 1.5) \cdot 10^{15} \sqrt{A}$$

where $k = 1.38 \times 10^{-23}$ Boltzmann constant, J/°K; $Z_1, Z_2$ are charge state of 1 and 2 particles respectively; $e = 1,6 \times 10^{-19}$ C is charge of electron; $r_o$ is radius of nuclear force, m; $A$ is number of element; $F$ is force, N; $E$ is energy, J; $Q$ is charge of particles.

For example, for reaction H+H (hydrogen, $Z_1 = Z_2 = 1$, $r_o \approx 2 \times 10^{-15}$ m) this energy is $\approx 0.7$ MeV or 0.35 MeV for every particle. The real energy is about 30 times less because some particles have more than average speed and there is a tunnel effect.

**2. Energy needed for ignition.** Fig. 8 [15] shows a magnitude $n\tau$ (analog of Lawson criterion) required for ignition.

Present-day industry produces powerful lasers:
- Carbon dioxide lasers emit up to 100 kW at 9.6 μm and 10.6 μm, and are used in industry for cutting and welding.
- Carbon monoxide lasers must be cooled, but can produce up to 500 kW.

Special laser and ICF reactors:

- NOVA (1999, USA). Laser 100 kJ (wavelenght λ=1054×10⁻⁹ m) and 40 kJ (wavelenght λ=351×10⁻⁹ m), power few tens of terawatts (1 TW = $10^{12}$ W), time of impulse (2 ÷ 4) ×10⁻⁹ s, 10-beams, matter is Nd:class.
- OMERA (1995, USA). 60-beam, neodyminm class laser, 30 kJ, power 60 TW.
- Z-machine (USA, under construction), power is up 350 TW. It can create currency impulses up to $20 \times 10^6$ A.
- NIF (USA). By 2005, the National Ignition Facility is worked on a system that, when complete, will contain 192-beam, 1.8-megajoule, 700-terawatt laser system adjoining a 10-meter-diameter target chamber.
- 1.25 PW - world's most powerful laser (claimed on 23 May 1996 by Lawrence Livermore Laboratory).

**3. Radiation energy from hot solid black body** is (Stefan-Boltzmann Law):
$$E = \sigma T^4, \qquad (2)$$
where $E$ is emitted energy, W/m²; $\sigma = 5.67 \times 10^{-8}$ - Stefan-Boltzmann constant, W/m² °K⁴; $T$ is temperature in °K.

**4. Wavelength** corresponded of maximum energy density (Wien's Law) is
$$\lambda_0 = \frac{b}{T}, \quad \omega = \frac{2\pi}{\lambda_0}, \qquad (3)$$
where $b = 2.8978 \times 10^{-3}$ is constant, m °K; $T$ is temperature, °K; $\omega$ is angle frequency of wave, rad/s.

**5. Pressure for one full reflection** is
$$F = 2E/c, \qquad (4)$$
where $F$ - pressure, N/m²; $c = 3 \times 10^8$ is light speed, m/s, $E$ is radiation power, W/m². If plasma does not reflect radiation the pressure equals
$$F = E/c. \qquad (5)$$

**6. Pressure for plasma multi-reflection** [8, 14] is
$$F = \frac{2E}{c}\left(\frac{2}{1-q}\right), \qquad (6)$$
where $q$ is plasma reflection coefficient. For example, if $q = 0.98$ the radiation pressure increases by 100 times. We neglect losses of prism reflection.

**7. The Bremsstrahlung (brake) loss** energy of plasma by radiation is ($T > 10^6$ °K)
$$P_{Br} = 5.34 \cdot 10^{-37} n_e^2 T^{0.5} Z_{eff}, \quad \text{where} \quad Z_{eff} = \sum (Z^2 n_z)/n_e, \qquad (7)$$
where $P_{Br}$ is power of Bremsstrahlung radiation, W/m³; $n_e$ is number of particles in m³; $T$ is a plasma temperature, KeV; $Z$ is charge state; $Z_{eff}$ is cross-section coefficient for multi-charges ions. For reactions H+D, D+T the $Z_{eff}$ equals 1.

Losses may be very high. For some reactions, they are more then useful nuclear energy and fusion nuclear reaction may be stopped. The Bremsstrahlung emission has continuous spectra.

**8. Electron frequency in plasma** is
$$\omega_{pe} = \left(\frac{4\pi n_e e^2}{m_e}\right)^{1/4}, \quad \text{or} \quad \omega_{pe} = 5.64 \times 10^4 (n_e)^{1/4}, \qquad (8)$$
in "cgs" units, or $\omega_{pe} = 56.4(n)^{1/4}$ in CI units

where $\omega_{pe}$ is electron frequency, rad/s; $n_e$ is electron density, [1/cm³]; $n$ is electron density, [1/m³]; $m_e = 9.11 \times 10^{-28}$ is mass of electron, g; $e = 1.6 \times 10^{-19}$ is electron charge, C.

The plasma is reflected an electromagnet radiation if frequency of electromagnet radiation is less then electron frequency in plasma, $\omega < \omega_{pe}$. That reflectivity is high. For $T > 15 \times 10^6$ °K it is more than silver and increases with plasma temperature as $T^{3/2}$. The frequency of laser beam and Bremsstrahlung emission are less then electron frequency in plasma.

**9. The deep of penetration** of outer radiation into plasma is

$$d_p = \frac{c}{\omega_{pe}} = 5.31 \cdot 10^5 n_e^{-1/2} \quad \text{[cm]} \tag{9}$$

For plasma density $n_e = 10^{22}$ 1/cm$^3$ $d_p = 5.31 \times 10^{-6}$ cm.

**10. The gas (plasma) dynamic pressure**, $p_k$, is

$$p_k = nk(T_e + T_i) \quad \text{if} \quad T_e = T_k \quad \text{then} \quad p_k = 2nkT, \tag{10}$$

where $k = 1.38 \times 10^{-23}$ is Boltzmann constant; $T_e$ is temperature of electrons, °K; $T_i$ is temperature of ions, °K. These temperatures may be different; $n$ is plasma density, 1/m$^3$; $p_k$ is plasma pressure, N/m$^2$.

**11. The gas (plasma) ion pressure**, $p$, is

$$p = \frac{2}{3} nkT, \tag{11}$$

Here $n$ is plasma density in 1/m$^3$.

**12. The magnetic $p_m$ and electrostatic pressure**, $p_s$, are

$$p_m = \frac{B^2}{2\mu_0}, \quad p_s = \frac{1}{2}\varepsilon_0 E_S^2, \tag{12}$$

where $B$ is electromagnetic induction, Tesla; $\mu_0 = 4\pi \times 10^{-7}$ electromagnetic constant; $\varepsilon_0 = 8.85 \times 10^{-12}$, F/m, is electrostatic constant; $E_S$ is electrostatic intensity, V/m.

**13. Ion thermal velocity** is

$$v_{Ti} = \left(\frac{kT_i}{m_i}\right)^{1/2} = 9.79 \times 10^5 \mu^{-1/2} T_i^{1/2} \quad \text{cm/s}, \tag{13}$$

where $\mu = m_i/m_p$, $m_i$ is mass of ion, kg; $m_p = 1.67 \times 10^{-27}$ is mass of proton, kg.

**14. Transverse Spitzer plasma resistance**

$$\eta_\perp = 1.03 \times 10^{-2} Z \ln \Lambda T^{-3/2}, \quad \Omega \text{ cm} \quad \text{or} \quad \rho \approx \frac{0.1 Z}{T^{3/2}} \quad \Omega \text{ cm}, \tag{14}$$

where $\ln \Lambda = 5 \div 15 \approx 10$ is Coulomb logarithm, $Z$ is charge state.

**15. Reaction rates** $\langle \sigma v \rangle$ (in cm$^3$ s$^{-1}$) averaged over Maxwellian distributions for low energy (T<25 keV) may be represent by

$$\begin{aligned}(\overline{\sigma v})_{DD} &= 2.33 \times 10^{-14} T^{-2/3} \exp(-18.76 T^{-1/3}) \quad \text{cm}^3\text{s}^{-1}, \\ (\overline{\sigma v})_{DT} &= 3.68 \times 10^{-12} T^{-2/3} \exp(-19.94 T^{-1/3}) \quad \text{cm}^3\text{s}^{-1},\end{aligned} \tag{15}$$

where T is measured in keV.

**16. The power density** released in the form of charged particles is

$$\begin{aligned}P_{DD} &= 3.3 \times 10^{-13} n_D^2 (\overline{\sigma v})_{DD}, \quad \text{W cm}^{-3} \\ P_{DT} &= 5.6 \times 10^{-13} n_D n_T (\overline{\sigma v})_{DT}, \quad \text{W cm}^{-3} \\ P_{DHe^3} &= 2.9 \times 10^{-12} n_D n_{He^3} (\overline{\sigma v})_{DHe^3}, \quad \text{W cm}^{-3}\end{aligned} \tag{16}$$

Here in $P_{DD}$ equation it is included D + T reaction.

## Results of computation

**1. Some thermonuclear reactions.** The primary nuclear reaction is D-D reaction that takes place when two nuclei of deuterium collide. Deuterium can be obtained from seawater, its abundance being about 0.0148% that of hydrogen, and used as a fuel resource, this amount can be regarded as almost inexhaustible.

The D-D reaction consists of the following two reactions:

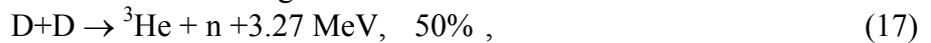
D+D → $^3$He + n +3.27 MeV,   50% ,   (17)
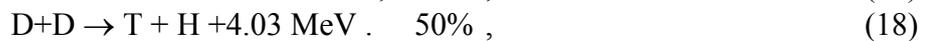
D+D → T + H +4.03 MeV .   50% ,   (18)

In reaction (17) an isotope of helium ($^3$He) and neutron (n) are produced by the collision of two deuterium nuclei (D). In reaction (2), a tritium (T) and a proton (H) are produced. The numbers on the right-side denote the kinetic energy released by the reaction, which can be calculated us follows: If we denote the mass defect of each particle in the unit of MeV (10$^6$ eV), we have D:

13.1359 Mev, He: 14.9313 MeV, and n: 8.0714 Mev ([16], p. 1295), so that the energy released by reaction (17) is

$$2 \times 13.1359 - (14.9313 + 8.0714) = 3.2691 = 3.27 \text{ MeV}.$$

For reaction (18), we can use for T: 14.9500 Mev and for H: 7.289 MeV.

The partition of the released energy from the reaction products can be estimated from energy and momentum conservation. Kinetic energy of D before collision is very small compared to the energy released by the reaction. We can ignore the initial kinetic energy and treat the deuterium nuclei as being at rest. Denoting the mass and speed of helium and n by $_1$ and $_2$ respectively, we have for reaction (17)

$$0.5 m_1 v_1^2 + 0.5 m_2 V_2^2 = E = 3.27 \quad \text{MeV}, \quad m_1 v_1 = m_2 v_2, \tag{19}$$

where in the second formula we assumed that He and n fly out in opposite directions. From these relations, we find

$$E_1 = \frac{1}{2} m_1 v_1^2 = \frac{E}{1 + m_1/m_2} = 0.82 \quad \text{MeV}, \quad E_2 = \frac{1}{2} m_2 v_2^2 = \frac{E}{1 + m_2/m_1} = 2.45 \quad \text{MeV}. \tag{20}$$

Obviously, the lighter particle acquires more energy than the heavier particle.

Current nuclear fusion research is focused on the D + T thermonuclear fusion reaction

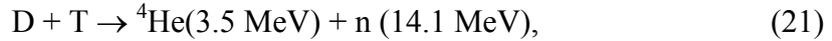

$$\text{D} + \text{T} \rightarrow {}^4\text{He}(3.5 \text{ MeV}) + \text{n} (14.1 \text{ MeV}), \tag{21}$$

Reaction (21) can occur in high-temperature deuterium-tritium plasma. Most energy released by the reaction is converted to the kinetic energy of the neutron. Since the neutron is not confined or reflected by a magnetic or electrostatic field it leaves, going outwards to surrounding space or hits the screen or vessel wall (or blanker) immediately after reaction. In last instance, the neutron kinetic energy is converted to heat. The heat is taken away from the screen by direct radiation or and indirect circulating coolant and can be used to run an electric generator. If we add $^6$Li inside the blanket, then tritium can be produced by reaction

$$\text{n} + {}^6\text{Li} \rightarrow {}^4\text{He} (2.1 \text{ MeV}) + \text{T} (2.7 \text{ MeV}) \tag{22}$$

and then used as the fuel. Another reaction product is the alpha particles $^4$He carrying 3.5 MeV which can be directed or confined by electro-magnetic field.

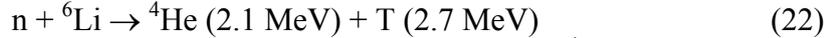

The reaction that produces only charged particles are best for the proposed propulsion system and generator. Unfortunately, these reactions are not great (see Table 1).

However, since the nuclei are positively charged, they must have enough energy to overcome the Coulomb repulsion between them, in order for a few of them to be able to combine. The required energy can be estimated by equation (1).

**2. Rocket Impulse.** When we know the energy of the thermonuclear particles, the particle speed can be calculated by equation

$$V_i = 1.384 \times 10^4 (E/N_i)^{0.5}, \tag{23}$$

where $E$ is particle energy in MeV, $N$ is number of nucleons in particle (in mass units, for example, $N = 1$ for proton, neutron, $N = 2$ for deuterium, $N = 3$ for tritium, $N = 4$ for helium).

Conventionally, we have two components on the equation's right-side having different mass and speed. The average efficiency particle speed $V$ (impulse) of thermonuclear reaction may be estimated by equation

$$V = \eta_1 \frac{N_1}{N} V_1 + \eta_2 \frac{N_2}{N} V_2. \tag{24}$$

where lower index "$_i$" is number of particle, $\eta$ is coefficient utilization of kinetic particle energy, $N$ is total number of nucleons in single reaction. For particles adsorbed by screen (fig.1a) $\eta = 0.25$, for particles reflected by reflector (fig.1b) $\eta = 1$.

When we use the radiation (photon) energy of one hot side of screen, the efficiency particle speed (24) has additional member

$$\Delta V = \eta_3 \frac{E_J}{m_p c N}, \tag{25}$$

where $m_p = 1.67495 \times 10^{-27}$ kg is mass of neutron, $c = 3 \times 10^8$ m/s is the light speed, $E_J = 1.6 \times 10^{-19} E$ energy of particle in J, $\eta_3 = 0.25$ is coefficient utilization of heat energy.

The apparatus speed is

$$V_m = -V \ln \frac{M_f}{M_0}, \qquad (26)$$

where $V_m$ is maximum speed of space apparatus, m/s; $M_f/M_0$ is ratio of a final mass (apparatus without thermonuclear fuel) to the initial apparatus mass.

Results of computation are presented in Table #1.

Table 1.

| Type of propulsion→<br><br>Thermonuclear reaction, MeV ↓ | AB-screen, Impulse from charged particles ×10⁶ m/s | AB-screen. Max speed of apparatus for $M_f/M_0$=0.1 Speed×10⁶m/s | AB-reflector. Impulse from charged particles ×10⁶ m/s | AB-reflector. Max speed of apparatus for $M_f/M_0$=0.1 Speed×10⁶m/s | Mass ratio $M_f/M_0$ for fuel=0 | + Photon. Add speed ×10⁶m/s |
|---|---|---|---|---|---|---|
| 1 | 2 | 3 | 4 | 5 | 6 | 7 |
| D+T→⁴He(3.5)+n(14.1) | 5.18 | 8.13 | 10.3 | 23.8 | 0.19 | 0.23 |
| T+T→⁴He(3.77)+2n(7.53) | 4.48 | 7.03 | 8.96 | 20.6 | 0.25 | 0.15 |
| D+³He→⁴He(3.6)+p(14.7) | 5.28 | 8.29 | 21.1 | 48.5 | 0.1 | - |
| D+⁶Li→2⁴He(22.4) | 5.8 | 9.11 | 23.6 | 54.3 | 0.1 | - |
| ³He+³He→⁴He(4.3)+2p(8.6) | 5 | 7.85 | 20 | 46 | 0.1 | - |
| ³He+⁶Li→2⁴He(1.9)+p(16.9) | 3 | 4.47 | 12 | 27.6 | 0.1 | - |
| p+¹¹B→3⁴He(8.7) | 2.95 | 4.63 | 11.78 | 27.1 | 0.1 | - |

Here are: D - deuterium, T - tritium, He - helium, Li - lithium, B - boron, n - neutron, p - proton.

The *first* column shows the thermonuclear reaction. In left side from pointer it is shown the components of thermonuclear reactor fuel. In the right-side it is shown the particles which appear in the reaction and kinetic energy every particle in MeV.

The *second* column shows the efficiency impulse (in m/s) computed by equation (18) for **AB-screen** engine.

The *third* column shows the maximum speed which apparatus (equipped with an engine screen) reaches a fuel mass ratio equal to 0.1 (equation (20)).

The *fourth* and *fifth* column shows the efficiency impulse and maximum apparatus speed for **AB-reflector** engine.

The *sixth* column shows the final mass of the space apparatus when the mass of initial fuel equals zero. In the first two reactions this ratio is not 0.1 because the neutrons are adsorbed by the screen. That decreases the apparatus speed, but the neutrons can be harnessed to get additional fuel and to sustain other useful and valuable thermonuclear reaction. All computation in Table 1 is made for mass ratio 0.1.

The last (*seventh*) column shows the additional speed from hot one-sided screen emitting photons. This additional speed is small.

The thrust of the spherical thruster-generator (fig.1*d*) for small angles may be computed as for AB-screen engine. Thrust is proportion the ratio of the open area to the full sphere surface.

The power of electric generator may be estimated by equation

$$W = 0.5 \times 10^{14} \eta \frac{N_p}{N} m_f E, \qquad (27)$$

where $W$ is power, W; $N_p$ is number of the charges (protons) nucleons; $N$ is total number of nucleons; $E$ is energy of charged particles, MeV; $m_f$ is fuel consumption, kg/s. $\eta$ is coefficient efficiency.

The trust propulsion is

$$T = m_f V, \qquad (28)$$

where $T$ is trust, N.

The relative mass of fuel converted to energy and thrust is

$$\Delta \overline{m} = \frac{E}{938 N}, \qquad (29)$$

where $E$ is reaction energy in MeV.

For example, let us take the reaction D+T and fuel consumption $m_f = 10^{-5}$ kg/s. Then $N_p/N = 4/5$, $\underline{E} = 3.5$ MeV and $W \approx 1.34 \times 10^6$ kW; thrust $T \approx 100$ N; the relative mass converted into energy is $3.75 \times 10^{-3}$. If the fuel consumption is $m_f = 10^{-2}$ kg/s, the thrust is $T \approx 10^5$ N. The energy is gigantic $W \approx 1.34 \times 10^9$ kW.

Table 1 shows that the offered thermonuclear AB-propulsions can accelerate the space apparatus up the speed $(20 \div 50) \times 10^6$ m/s (or up 1/6 of light speed) with a fuel ratio of $M_f/M_0 = 0.1$. The AB-propulsion is the most efficient of all thermonuclear propulsions, capable of reaching the theoretic maximum impulse from currently known thermonuclear propulsions and known thermonuclear reactions.

Please note that the reaction that produces only charged particles is more efficient than reactions producing neutrons and charged particles (two in the first lines of Table 1). The neutrons accept a lot of a common energy, but this energy does not produce any thrust. Converting this energy into photons (column 7 of table 1) is also ineffective. The neutrons do not leave the space apparatus, increasing its final launch and travel weight (column 6) and decreasing the final apparatus speed. They may be used for next reaction (see (22) and below), but technical realization of such reaction is decidedly complex and presently unproductive. See some of these reactions below. Unfortunately, most neutron reactions are exoteric.

3. **The thermonuclear reactions used the slow neutrons:**

Table 2

| Reaction | Energy, MeV | Cross section, burns |
|---|---|---|
| $^3$He + n → $^3$H+p | +0.764 | 5400 |
| $^6$Li + n → $^3$H+α | +4.785 | 945 |
| $^7$Be + n → $^7$Li+p | +1.65 | 51000 |
| $^{10}$B + n → $^7$Li+α | +2.791 | 3837 |
| $^{14}$N + n → $^{14}$C+p | +0.626 | 1.75 |
| $^{17}$O + n → $^{14}$C+α | +1.72 | 0.5 |
| $^{33}$S + n → $^{33}$P+p | +0.75 | 0.002 |
| $^{35}$Cl +n → $^{35}$H+p | +0.62 | 0.3 |

The spherical AB-engine can produce much electrical energy, but conversion of this energy into vehicular thrust by common electric (ion) propulsion is inefficient in comparison with the offered AB-engine.

## Discussion

The potential space traveling apparatus speed 1/6 of light speed is maximum velocity predicted by thermonuclear AB-propulsion. That speed allows Mars to be a destination in minutes (or some days when apparatus has limited acceleration); that very high speed allows short period trips throughout our Solar system. However, it is not sufficient for easy interstellar space trips. The nearest star system is located at a distance of 3 - 5 light-years. That means the trip requires a minimum of 40 - 60 years. But required fuel ratio $M_f/M_0$ is very high: acceleration and braking of moving apparatus needs 4-stage rocket having the ratio for every stage $M_f/M_0 = 0.1$. The total weight fuel ratio will be $M_f/M_0 = 10^{-4}$. If useful weight is 10 tonnes, the starting rocket mass is $M_0/M_f = 10^4$ tonnes. The relative mass of thermonuclear reaction converted into energy (and thrust) is only $0.3 \div 0.4\%$ of total fuel mass. The author's research so far shows that the magnet cannot adsorb the big amount of interstellar matter in the high apparatus speed mode; consequently, the envisioned apparatus must take fuel for the entire trip.

Human interstellar flight is very expensive and complex. We can develop long-distance communication system and send, instead, E-men [18, 19] or artificial intelligent robot.

However, only an annihilation reaction can efficiently solve the interstellar trip macro-problem. Otherwise, new physics discoveries that allow such trips is required.

## Conclusion

The author suggests the simplest maximally efficient thermonuclear AB-propulsion (and electric generators) based in the early offered size-minimized Micro-AB-thermonuclear reactor [15]. These engines directly convert high-speed charged particles produced in thermonuclear reactor into vehicular thrust or onboard vehicle electricity resource. Offered propulsion system allows travel to any of our Solar System's planets in a short time as well as trips to the nearest stars by E-being or intellectual robot in during a single human life [18 ]- [19].

## Acknowledgement

The author wishes to acknowledge Richard Cathcart for correcting the English and other useful advices.